\documentclass[prl,floatfix,superscriptaddress,twocolumn,showpacs,amsmath,amssymb]{revtex4-1}

\usepackage{graphicx}% Include figure files
\usepackage{dcolumn}% Align table columns on decimal point
\usepackage{bm}% bold math
\usepackage{lipsum}
\usepackage{comment}
\newcommand{\mv}[1]{\ensuremath{\mathbf{#1}}} % for vectors
 
\newcommand{\avg}[1]{\left \langle #1 \right \rangle} % for average
\newcommand{\pd}[2]{\frac{\partial #1}{\partial #2}} % for partial derivatives
\newcommand{\pdd}[2]{\frac{\partial^2 #1}{\partial #2^2}} % for double partial derivatives

\usepackage[colorlinks]{hyperref}  
\usepackage{cleveref}
\hypersetup{linkcolor=blue,citecolor=blue,urlcolor=blue}

\usepackage{titlesec}
\titleformat{\section}
  {\centering\normalfont\fontsize{11}{15}\bfseries}{\thesection}{1em}{}

\usepackage[toc]{appendix}

\crefname{figure}{Fig.}{Figs.}
\Crefname{figure}{Figures}{Figures}
\crefname{equation}{Eq.}{Eqs.}

\begin{document}

\setcounter{page}{1} %first page number

\title{Glassy Dynamics in Chiral Fluids}

\author{Vincent E. Debets}
\affiliation{Department of Applied Physics, Eindhoven University of Technology, P.O. Box 513,
5600 MB Eindhoven, The Netherlands}
%\affiliation{Institute for Complex Molecular Systems, Eindhoven University of Technology, P.O. Box 513, 5600 MB Eindhoven, The Netherlands}

\author{Hartmut L\"owen}
\affiliation{Institut für Theoretische Physik II: Weiche Materie, Heinrich-Heine-Universität Düsseldorf, D-40225 Düsseldorf, Germany\\}

\author{Liesbeth M.C. Janssen}
\affiliation{Department of Applied Physics, Eindhoven University of Technology, P.O. Box 513,
5600 MB Eindhoven, The Netherlands}
%\affiliation{Institute for Complex Molecular Systems, Eindhoven University of Technology, P.O. Box 513, 5600 MB Eindhoven, The Netherlands}
\email{l.m.c.janssen@tue.nl}

\begin{abstract}%
{\noindent Chiral active matter is enjoying a rapid increase of interest, spurred by the rich variety of asymmetries that can be attained in e.g.\ the shape or self-propulsion mechanism of active particles. Though this has already led to the observance of so-called chiral crystals, active chiral glasses remain largely unexplored. A possible reason for this could be the naive expectation that interactions dominate the glassy dynamics and the details of the active motion become increasingly less relevant. Here we show that quite the opposite is true by studying the glassy dynamics of interacting chiral active Brownian particles (cABPs).
We demonstrate that when our chiral fluid is pushed to glassy conditions, it exhibits highly nontrivial dynamics, especially compared to a standard linear active fluid such as common ABPs. Despite the added complexity, we are still able to present a full rationalization for all identified dynamical regimes. Most notably, we introduce a new 'hammering' mechanism, unique to rapidly spinning particles in high-density conditions, that can fluidize a chiral active solid.
%and is capable of inducing remarkably strong reentrant behavior.
%We therefore conclude that chirality gives rise to unexpected and new dynamical phenomena, even in the glassy regime.
%where it is usually argued that interactions dominate the dynamics. 
}
\end{abstract}

\maketitle %%The above information typeset through this command

%Mention some of the interesting non-eq features of active matter (like MIPS etc)
%Introduce active matter, followed by chiral fluid which converges on the interplay between persistence time and spinning frequency

\textit{Introduction.---} Inspired by its omnipresence in biology, as well as its growing relevance in condensed matter and materials science, active matter
%which usually consists of particles capable of autonomous motion through the stable consumption of energy, 
has proven to be one of the prevailing subjects in biological and soft matter physics% throughout the previous decade
~\cite{Bechinger2016,Ramaswamy2010,Marchetti2013rev}. Active or self-propelled particle systems  are intrinsically far from equilibrium, giving rise to a myriad of surprising features that are inaccessible to conventional passive matter. Well-known examples include motility induced phase separation (MIPS)~\cite{Buttinoni2013,Ginot2018,Palacci2013,Linden2019}, accumulation around repulsive obstacles~\cite{Berke2008}, spontaneous velocity alignment~\cite{Deseigne2010}, and active turbulence~\cite{Giomi2015,Alert2022rev}. Interestingly, so-called linear swimmer models such as active Brownian particles (ABPs)~\cite{Fily2012,Bialke2012,Redner2013,Caporusso2020,Ni2015,Caprini2020}, active Ornstein Uhlenbeck particles (AOUPs)~\cite{Maggi2015}, and run-and-tumble particles (RTPs)~\cite{Tailleur2008,Cates2013} have already been remarkably successful in theoretically describing a significant number of these non-equilibrium features. Members of this class of particles are typically endowed with a constant (average) self-propulsion whose direction changes randomly via some form of rotational diffusion (often thermal fluctuations). However, due to for instance an asymmetric shape~\cite{Kummel2013,Patra2022,Zhang2020}, mass distribution~\cite{Campbell2017}, or self-propulsion mechanism~\cite{Archer2015,Lauga2006}, active particles also frequently self-rotate which is not included in the aformentioned models. This leads to chiral-symmetry breaking of the corresponding active motion and, at small enough densities, circular (2D) or helical trajectories (3D). A collection of these spinning particles is usually referred to as an active chiral fluid and has been shown to exhibit many interesting collective phenomena in both simulations and experiments~\cite{Huang2020,Ni_cABP_2022,Ni2019_science_adv,Liebchen2017,Chen2017,Tan2022,Zhang2020,Scholz2021,Soni2019,Han2021,Reichhardt2019,Liao2018,Liao2021}. 
%For simple model systems this results from the nontrivial interplay between two competing timescales, i.e., the persistence time (or inverse rotational diffusion coefficient) and the spinning frequency~\cite{Ni_cABP_2022,Ni2019_science_adv}. 
Understanding the influence of chirality on active matter is therefore enjoying growing attention~\cite{Bowick2022,Liebchen2022review}, but at the same time requires more involved modelling efforts to fully comprehend.

\begin{figure}[ht!]
    \centering
    \includegraphics [width=0.5\textwidth] {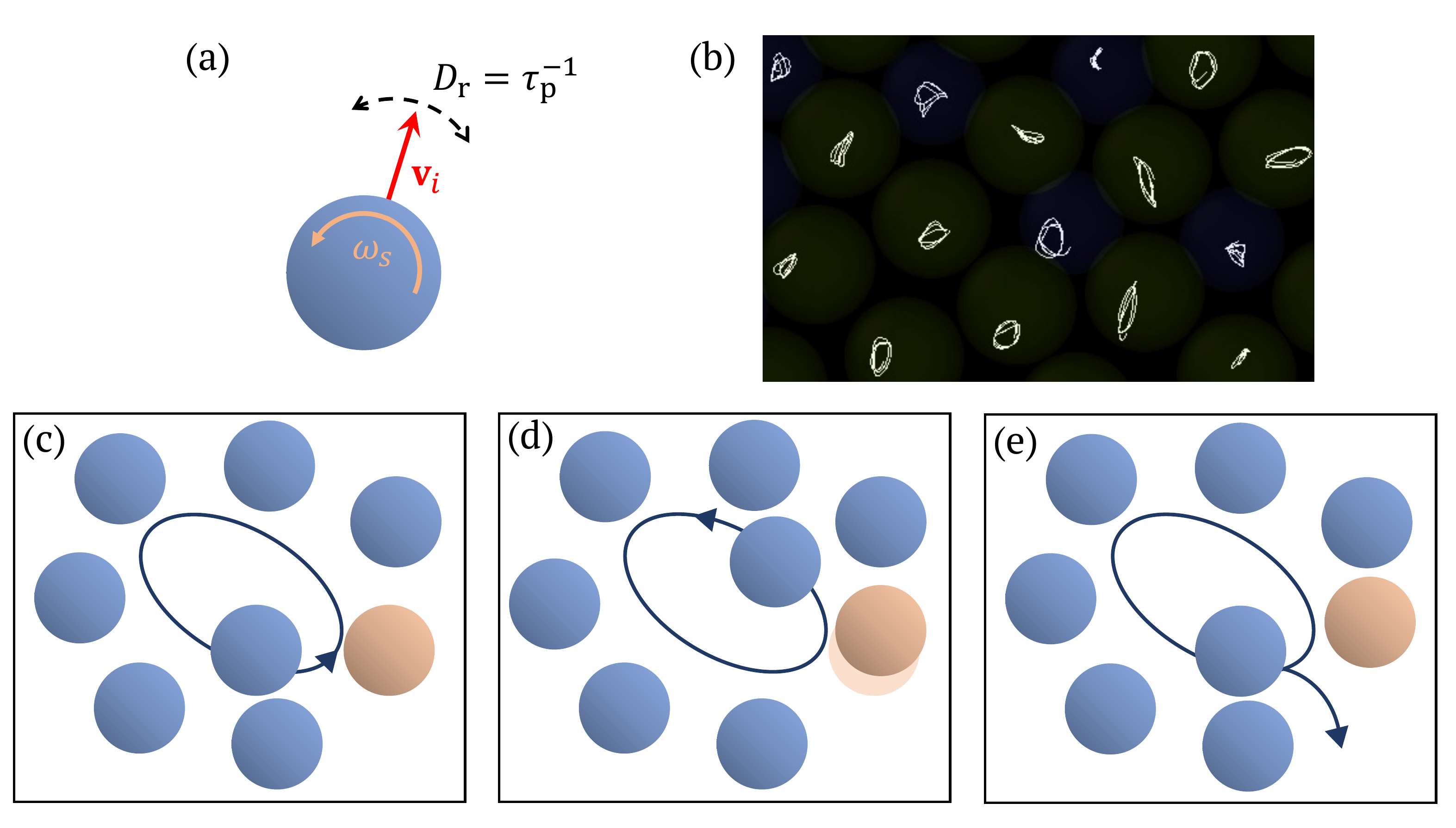} 
    \caption{(a) Visualization of a chiral active Brownian particle (cABP). (b) Example short-time trajectories (total time is equal to three spinning periods) of cABPs at large spinning frequency and persistence exhibiting the 'hammering' effect by undergoing circular motion inside their cage of surrounding particles. (c-e) Schematic depiction of the 'hammering' effect. (c-d) For large enough persistence and spinning frequency, particles undergo back-and-forth motion inside their cage and systematically collide with the same particle whose motion is slightly altered by the collision. (e) After repeated collisions the cage of a particle is sufficiently remodelled such that the particle can break out and migrate through the material. 
    }
    \label{Fig1}
\end{figure}

Initial chiral active matter studies have focused primarily on the low to moderate density regime~\cite{Lauga2006,Teeffelen2008,Ohta2009,DiLeonardo2011,Kummel2013}, but interest is now increasingly shifting towards high densities. This has already yielded several seminal works in the context of so-called chiral crystals~\cite{Petroff2015,Yan2015,Huang2020,Tan2022}. At the same time, their disordered counterpart, i.e., an active chiral glass, still remains largely unexplored. A possible reason for this could be that one naively expects active motion, at least to a large degree, to be impeded by interactions. As a result, the specific details of the active motion, whether chiral or nonchiral, should become increasingly less relevant upon approaching dynamical arrest. Here we demonstrate that in fact quite the opposite is true and that chiral active motion can certainly influence glassy dynamics in highly surprising ways. We, for the first time, delve into the unique physics that emerges when a chiral fluid ventures into the glassy regime. Most notably, we introduce a new 'hammering' mechanism (see~\cref{Fig1}), unique to rapidly spinning particles in high-density conditions, that can fluidize a chiral active solid. 

In short, we explore the dynamics of interacting chiral active Brownian particles (cABPs)~\cite{Teeffelen2008,Ni_cABP_2022} and show that when pushed to glassy conditions our chiral fluid exhibits highly nontrivial dynamics, particularly compared to standard linear active glassy matter (that is, conventional ABPs), which has already been extensively studied in theory~\cite{Voigtmann2017,SzamelABP2019,SzamelAOUP2015,SzamelAOUP2016,FengHou2017,Berthier2013activeglass,Reichert2020modecoupling,Reichert2020tracer,Reichert2021rev,Nandi2018} and simulation~\cite{BerthierABP2014,DijkstraABP2013,BerthierAOUP2017,Flenner2020,FlennerAOUP2016,Henkes2011active,Sollich2020,janzen2021aging,Bi2016cell,paoluzzi2022,Debets2021cage,Keta2022,Debets2022soft}. Despite the added complexity, we are still able to present a full rationalization for all identified dynamical regimes, including the emergence of a complex reentrant behavior which we explain by invoking the aforementioned 'hammering' mechanism. %Most notably, we introduce a new 'hammering' mechanism (see~\cref{Fig1}), unique to rapidly spinning particles in high-density conditions, that can fluidize a chiral active solid.  %inducing remarkably strong reentrant behavior which becomes more dramatic for faster spinning particles.

\textit{Simulation Details.---} As our model chiral fluid we consider a two-dimensional (2D) Kob-Andersen mixture which consists of $N_{\mathrm{A}}=650$ and $N_{\mathrm{B}}=350$ self-propelling quasihard disks of type A and B, respectively. We assume that the self-propulsion dominates over thermal fluctuations so that we can neglect passive diffusion and the equation of motion for the position $\mv{r}_{i}$ of each particle $i$ is given by~\cite{SzamelAOUP2016}
\begin{equation}\label{eom_r}
    \dot{\mv{r}}_{i} = \zeta^{-1} \mv{F}_{i} + \mv{v}_{i}.
\end{equation}
Here, $\zeta$ represents the friction constant and $\mv{v}_{i}$ the self-propulsion velocity acting on particle $i$. The interaction force $\mv{F}_{i}=-\sum_{j \neq i} \nabla_{i} V_{\alpha\beta}(r_{ij})$ is obtained from a quasihard sphere power law potential $V_{\alpha\beta}(r)= 4\epsilon_{\alpha\beta}\left( \frac{\sigma_{\alpha\beta}}{r}\right)^{36}$ \cite{Weysser2010structural,Lange2009} and the interaction parameters, i.e., $\epsilon_{\mathrm{AA}}=1,\  \epsilon_{\mathrm{AB}}=1.5,\  \epsilon_{\mathrm{BB}}=0.5,\  \sigma_{\mathrm{AA}}=1,\ \sigma_{\mathrm{AB}}=0.8,\  \sigma_{\mathrm{BB}}=0.88$, are, in combination with setting $\zeta=1$, chosen to frustrate crystallization and allow for glassy behavior \cite{Kob1994,Michele2004}. The choice of parameters also implies that we use reduced units where $\sigma_{\mathrm{AA}}$, $\epsilon_{\mathrm{AA}}$, $\epsilon_{\mathrm{AA}}/k_{\mathrm{B}}$, and $\zeta\sigma^{2}_{\mathrm{AA}}/\epsilon_{\mathrm{AA}}$ represent the units of length, energy, temperature, and time respectively~\cite{Flenner2005}.  For the self-propulsion of each particle we employ the cABP scheme~\cite{Teeffelen2008,Ni_cABP_2022}. That is, the magnitude of the self-propulsion or active speed $v_{0}$ is assumed to remain constant in time $t$ so that $\mv{v}_{i}=v_{0}\mv{e}_{i}=v_{0}[\cos(\theta_{i}),\sin(\theta_{i})]$, while the orientation angle of the active velocity $\theta_{i}$ evolves in time according to
\begin{equation}
    \dot{\theta}_{i} = \chi_{i} + \omega_{\mathrm{s}},
\end{equation}
with $\omega_{\mathrm{s}}$ a constant spinning frequency, $\chi_{i}$ a Gaussian noise process with zero mean and variance $\avg{\chi_{i}(t)\chi_{j}(t^{\prime})}_{\mathrm{noise}}=2D_{\mathrm{r}}\delta_{ij}\delta(t-t^{\prime})$, and $D_{\mathrm{r}}$ the rotational diffusion coefficient. As our control parameters we take $\omega_{\mathrm{s}}$, the persistence time $\tau_{\mathrm{p}}=D^{-1}_{\mathrm{r}}$, and a so-called spinning temperature $T_{\omega_{\mathrm{s}}}=v_{0}^{2}/2\omega_{s}$ which represents (up to a prefactor $4\pi\zeta$) a measure for the amount of energy that is dissipated by a single cABP during one circle motion.  
%This active matter model is thus described by three separate variables, i.e., $v_{0}$, $\omega_{\mathrm{s}}$, and $D_{\mathrm{r}}$ (or equivalently the persistence time $\tau_{\mathrm{p}}=D^{-1}_{\mathrm{r}}$), which will serve as our control parameters to study the active glassy behavior.

Simulations are performed by solving the overdamped equation of motion [\cref{eom_r}] via a forward Euler scheme using LAMMPS~\cite{Lammps}. We set the cutoff radius at $r_{\mathrm{c}}=2.5\sigma_{\alpha\beta}$, fix the size of the periodic square simulation box to ensure that the number density equals $\rho=1.2$, run the system sufficiently long (typically between $500$ and $10000$ time units) to prevent aging, and afterwards track the particles over time for at least twice the initialization time. 
%All results are presented in reduced units where $\sigma_{\mathrm{AA}}$, $\epsilon_{\mathrm{AA}}$, $\epsilon_{\mathrm{AA}}/k_{\mathrm{B}}$, and $\zeta\sigma^{2}_{\mathrm{AA}}/\epsilon_{\mathrm{AA}}$ represent the units of length, energy, temperature, and time respectively~\cite{Flenner2005}. 
To correct for diffusive center-of-mass motion all particle positions are retrieved relative to the momentary center of mass~\cite{Flenner2005}.

\begin{figure}[ht!]
    \centering
    \includegraphics [width=0.4\textwidth]
    %{D_tau_ws200_Tws4_p36_theory_sim_regimes.pdf} 
    {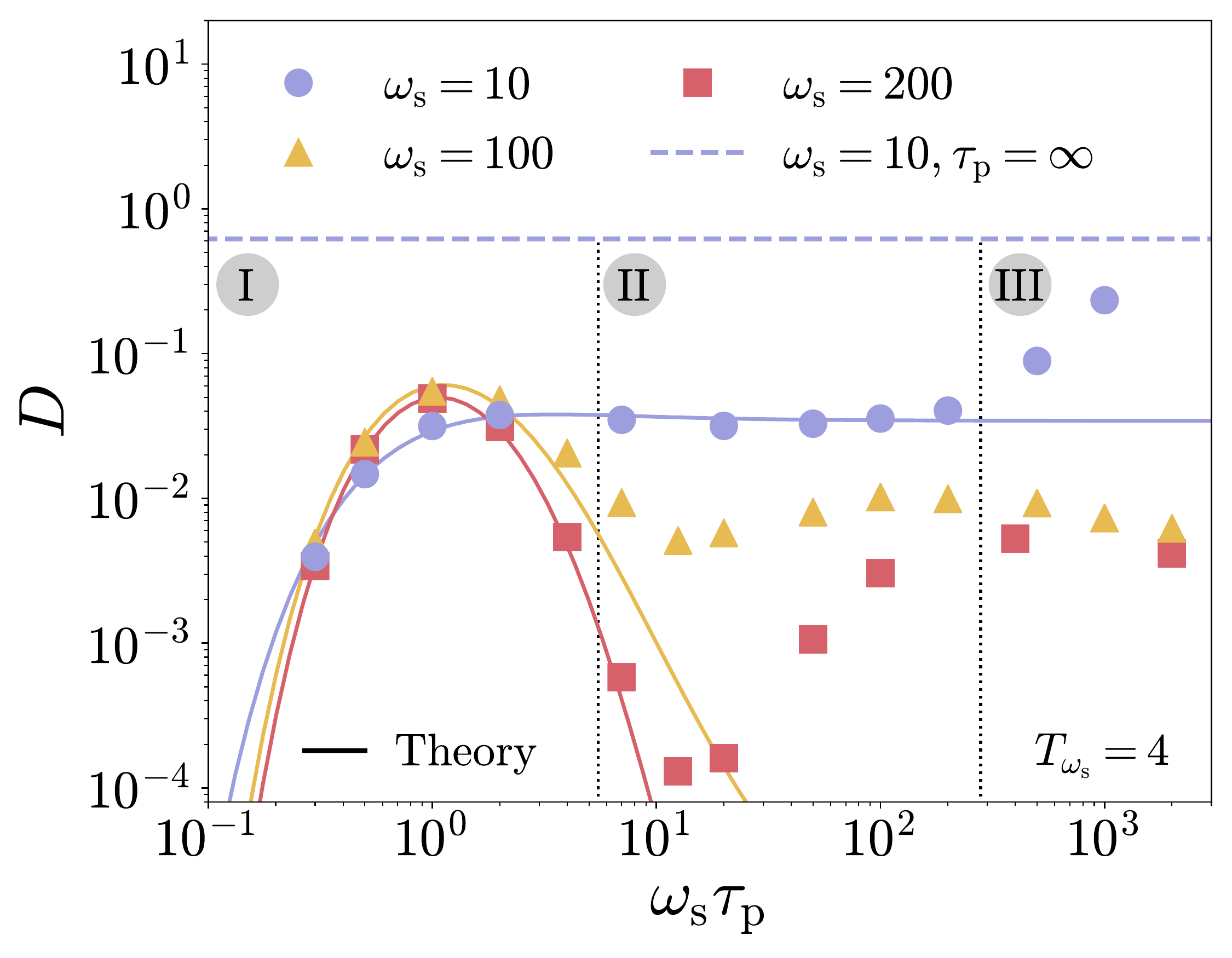}
    \caption{The long-time diffusion coefficient 
    $D$ as a function of the normalized persistence time $\omega_{\mathrm{s}} \tau_{\mathrm{p}}$ for several set values of $\omega_{\mathrm{s}}$ keeping $T_{\omega_{\mathrm{s}}}=4$ fixed. The resulting dynamics show highly non-trivial behavior that can be characterized by a nonmonotonic (I), reentrant (II), and large persistence (III) regime. The dashed line indicates the infinite persistence limit which is only nonzero for $\omega_{\mathrm{s}}=10$. 
    }
    \label{Fig2}
\end{figure}

\textit{Nonmonotonic Dynamics.---}  We are primarily interested in characterizing how the interplay between rotational diffusion and spinning motion influences the active glassy dynamics. Therefore, we have calculated the long-time diffusion coefficient $D=\lim_{t\to\infty} \avg{\Delta\mv{r}_{i}^{2}(t)}/4t$ of our chiral fluid for several set spinning frequencies $\omega_{\mathrm{s}}=10, 100, 200$ (keeping a fixed value $T_{\omega_{\mathrm{s}}}=4$ to ensure moderately supercooled behavior), while varying the persistence time.
The results are plotted as a function of $\omega_{\mathrm{s}}\tau_{\mathrm{p}}$ in~\cref{Fig2} and show remarkably rich dynamics. In particular, we find initial nonmonotonic behavior with a maximum at $\omega_{\mathrm{s}}\tau_{\mathrm{p}}\sim 1$. This is followed by a form of reentrant behavior which becomes much more pronounced for higher spinning frequencies. For example, at $\omega_{\mathrm{s}}=200$ the dynamics reaches a minimum with $D\sim10^{-4}$, which is practically a frozen system like a glass, that is seen to increase with orders of magnitude. Finally, in the limit of large persistence different asymptotic values ranging from significantly enhanced to zero dynamics are reached. To contrast these complex dynamics, we emphasize that a glassy liquid of standard ABPs at constant active speed $v_{0}$ would only show a monotonic enhancement of the dynamics for increasing persistence time~\cite{DijkstraABP2013,Voigtmann2017}. Thus, at large densities chirality has a highly nontrivial impact on active particle motion.

Moreover, we have verified that the same qualitative behavior is observed for both a different model glassformer and a different set of parameters where we have fixed the active speed $v_{0}$ instead of the spinning temperature $T_{\omega_{\mathrm{s}}}$ (see Figs.~S1~and~S2). We also mention that the nontrivial change of the dynamics and in particular the reentrant behavior are equally visible in the structure factor, self-intermediate scattering function, and the non-Gaussian parameter (see Figs.~S3,~S4,~and~S5). The latter being a measure for dynamical heterogeneity. 

\textit{cABP in a Harmonic Trap.---} Our aim now is to better understand the complex dynamics, which, for convenience, we will separate in three distinct regimes (see roman numerals in~\cref{Fig2}). 
%Given the absence of spatial velocity correlations in the limit of weak persistence (regime I), 
We first turn our attention towards regime I. Here, the persistence of particles is still relatively weak and we therefore expect that especially in this regime the local environment of particles (or their cage) acts primarily as an effective confining potential. This then motivates a comparison of our simulation results to those of a single cABP in a harmonic trap of the form $U(r)=\kappa r^{2}/2$, with $r$ the radial distance inside the trap and $\kappa$ its strength. In particular, we have analytically derived an expression for the long-time limit of the MSD of such a trapped particle (see SI for details). This yields
\begin{equation}\label{MSD_longtime3}
    \delta \equiv \lim_{t\to\infty} \avg{\Delta\mv{r}^{2}(t)} =\frac{v_{0}^{2}(1+\frac{\omega_{\mathrm{k}}}{\omega_{\mathrm{s}}\tau_{\mathrm{p}}})}{k^{2}[\omega_{k}^{2}+(1+\frac{\omega_{k}}{\omega_{\mathrm{s}}\tau_{\mathrm{p}}})^{2}]},
\end{equation}
where $k=\kappa/\zeta$, and we have introduced the dimensionless spinning frequencies $\omega_{k}=\omega_{\mathrm{s}}/k$ and $\omega_{\mathrm{s}}\tau_{\mathrm{p}}$. Due to the high-density (or glassy) conditions, we can then postulate that a particle escapes its trap when it reaches a distance equal to its diameter, that is, when $r=\sigma_{\mathrm{AA}}$. Assuming a Kramers-like process~\cite{Gardiner1985,Gov2020trap}, the corresponding average escape time is given by
\begin{equation}
    t_{\mathrm{esc}}=t_{0}e^{\frac{U(\sigma_{\mathrm{AA}})}{\avg{U}}}=t_{0}e^{\frac{\sigma_{\mathrm{AA}}^{2}}{\delta}},
\end{equation}
with $t_{0}$ a constant prefactor and we have used that the average potential energy of the particle is equal to $\avg{U}=\kappa \delta/2$~\cite{Gov2020trap}. Moreover, we assume that after each escape the particle falls into a new trap with the same properties. In other words, the particle diffuses through space by hopping from trap to trap (or equivalently cage to cage). This allows us to quantitatively estimate the long-time diffusion coefficient as $D\approx\sigma^{2}_{\mathrm{AA}}/4t_{\mathrm{esc}}$, which can be compared to our simulation results. Note that the qualitative behavior of our theoretically predicted $D$ is thus fully determined by a single fit parameter $k$, while the absolute scale is set by $t_{0}$. 

The resulting theoretical predictions (fitted on the first five data points) are shown as straight lines in~\cref{Fig2}. Remarkably, we find almost quantitative agreement in regime I and approximately the same fit value of $k\sim10$ for all our settings (the latter is consistent with the fact that we do not change the density or interaction potential which supposedly determine this factor). This demonstrates that the interplay between rotational diffusion and spinning motion in the small persistence regime are well captured by our simple single particle model.

\textit{Collective Motion.---} Inspired by previous work in literature~\cite{Ni2019_science_adv} we now proceed to regime III. Here, we observe a sudden increase of the dynamics towards the infinite persistence limit for relatively small spinning frequencies ($\omega_{\mathrm{s}}=10$). For larger spinning frequencies, we instead see $D$ decreasing and probably moving towards 
the infinite persistence limit of $D=0$ (see Fig.~S6). Thus, for $\tau_{\mathrm{p}}\rightarrow\infty$ there exists a transition from a so-called active ($D>0$) to an absorbing ($D=0$) state upon increasing $\omega_{\mathrm{s}}$. This behavior is fully consistent with previous work conducted at lower densities~\cite{Ni2019_science_adv}. Our work shows that this phenomenology is retained in the high density or glassy regime.
%A followup on the details of this transition would be very interesting, but for now is left for future work.

\begin{figure}[ht!]
    \centering
    \vspace{0.0cm}
    \includegraphics [width=0.44\textwidth] 
    %[width=9cm,height=4.5cm] 
    {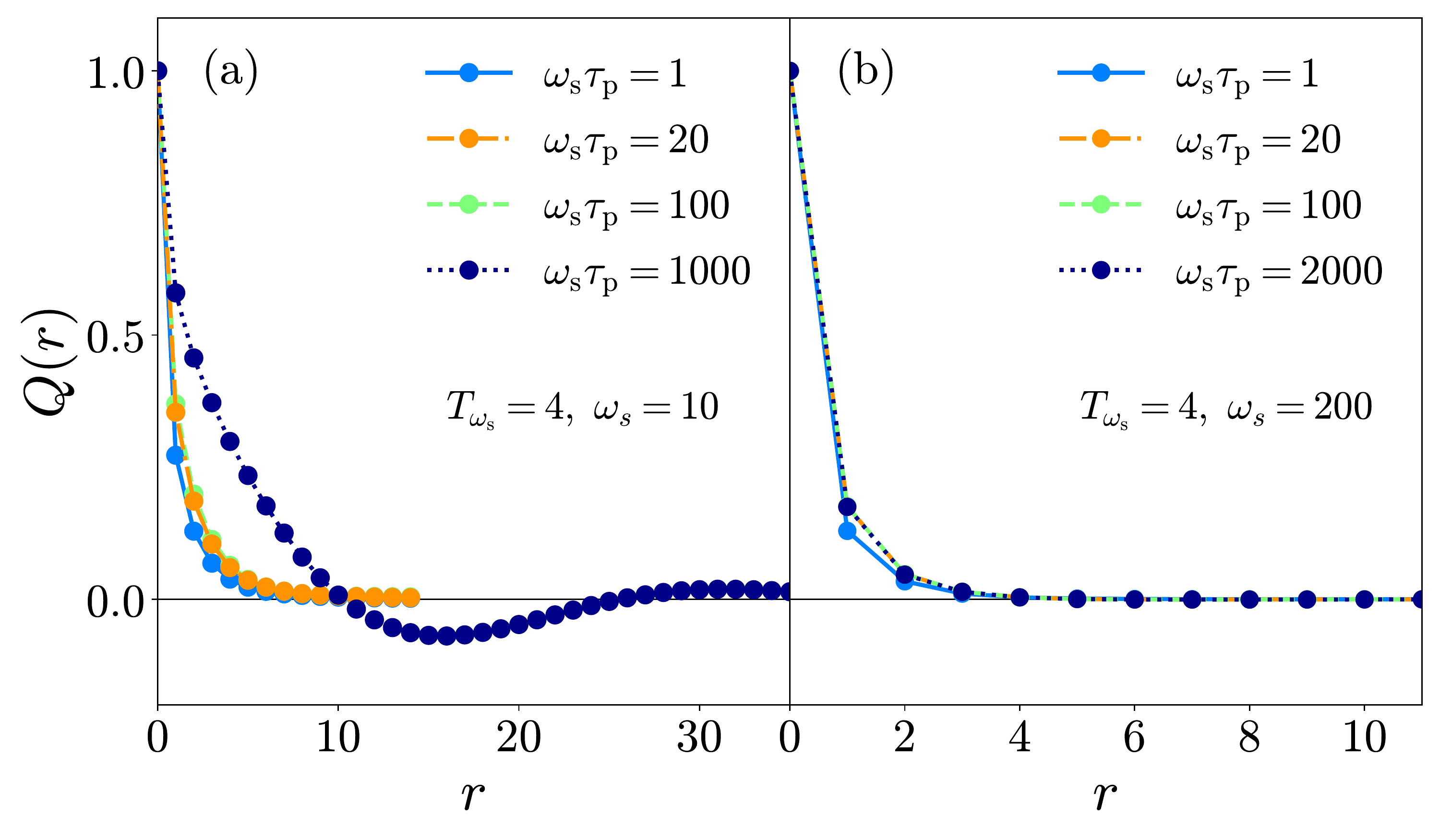} 
    \caption{The spatial velocity correlation function $Q(r)$ as a function of the distance $r$. (a) At small enough spinning frequencies ($\omega_{\mathrm{s}}=10$) we observe a sudden increase of spatial velocity correlations in the limit of large persistence ($\omega_{\mathrm{s}}\tau_{\mathrm{p}}=1000$). This increase is accompanied by a similar increase of the dynamics (see \cref{Fig2}) and a negative peak indicating vortex-like behavior. (b) In comparison, when particles spin too rapidly ($\omega_{\mathrm{s}}=200$) velocity correlations remain short-ranged and almost independent of persistence. }
    \label{Fig3}
\end{figure}

To explain why at small enough $\omega_{\mathrm{s}}$, that is, $\omega_{\mathrm{s}}=10$, the dynamics increases significantly in regime III, we  employ a spatial velocity correlation function $Q(r)$ (see~\cite{Caprini2020} for a precise definition). 
This function measures how correlated the velocities $\dot{\mv{r}}_{i}$ of different particles $i$ are over a distance $r$. It thus serves as a proxy for local velocity alignment and cooperative motion ($Q(r)=1,0,-1$ for perfect, no, and anti velocity alignment respectively). We have plotted $Q(r)$ for $\omega_{\mathrm{s}}=10$ and several values of $\omega_{\mathrm{s}}\tau_{\mathrm{p}}$ in~\cref{Fig3}a. In almost all cases we see a similar rapid decay to zero implying that there exists little velocity alignment and cooperative motion is absent. However, at exactly the same point where the dynamics has increased in regime III, i.e., $\omega_{\mathrm{s}}\tau_{\mathrm{p}}=1000$, we find that the decay of $Q(r)$ suddenly becomes much more long-ranged and even develops a negative peak (both features have been checked for finite-size effects). We interpret this as the particle motion becoming more collective and vortex-like which explains why its overall diffusion is enhanced. Moreover, note that this collective motion is only able to emerge when the amount of rotational diffusion is small enough, that is, when we are at a large enough $\tau_{\mathrm{p}}$. 

In comparison, for $\omega_{\mathrm{s}}=200$ we find almost no spatial velocity correlations for any value of $\omega_{\mathrm{s}}\tau_{\mathrm{p}}$ (see \cref{Fig3}b), which implies that no cooperative motion takes places. We expect that this is caused by particles spinning too rapidly which prevents them from inducing any collective motion, even in the absence of rotational diffusion ($\tau_{\mathrm{p}}\rightarrow\infty$). Ultimately, this should lead to particles becoming trapped in circular/elliptical trajectories inside their cage, thus explaining why $D$ goes to zero. 

%(which is diffusive due to particle-particle collisions)
%(where particles only move in circles/ellipses)

%which is illustrated by the large-time behavior of the mean square displacement at infinite persistence (see SI). For $\omega_{\mathrm{s}}=100.0, 200.0$ it oscillates around a plateau value (circular/elliptical motion), while for $\omega_{\mathrm{s}}=10.0$ it grows linearly in time (diffusive motion). In regime III of \cref{Fig2} we thus see $D$ move towards these infinite persistence limits.

\textit{Hammering Dynamics.---} We finalize our discussion by considering the intermediate persistence regime II. Interestingly, in this regime the agreement between theory and simulation only remains intact for a relatively small spinning frequency ($\omega_{\mathrm{s}}=10$, see~\cref{Fig2}). In comparison, for larger spinning frequencies a competing mechanism emerges which is able to increase the long-time dynamics from an almost glassy or dynamically arrested state ($D\sim10^{-4}$) with multiple orders of magnitude. We have checked that this reentrant behavior becomes even more extreme for larger values of $\omega_{\mathrm{s}}$. The key question therefore is, what causes such a dramatic increase of the dynamics if it is not cooperative diffusion (since there are almost no spatial velocity correlations, see \cref{Fig3}b). To answer this we propose a new 'hammering' mechanism that is distinct for rapidly spinning chiral particles at large densities (see \cref{Fig1}c-e for a schematic depiction). In short, for large enough persistence and spinning frequency, particles are expected to undergo long periods of uninterrupted back-and-forth motion inside their cage. During this they systematically collide with the same particle whose motion is slightly altered by each collision. After repeated collisions the cage of a particle is sufficiently remodelled such that the particle can break out and migrate through the material which should lead to faster dynamics.  

\begin{figure}[ht!]
    \centering
    \hspace{-0.0cm}
    \includegraphics [width=0.46\textwidth] 
    %[width=8.8cm,height=4.5cm] 
    {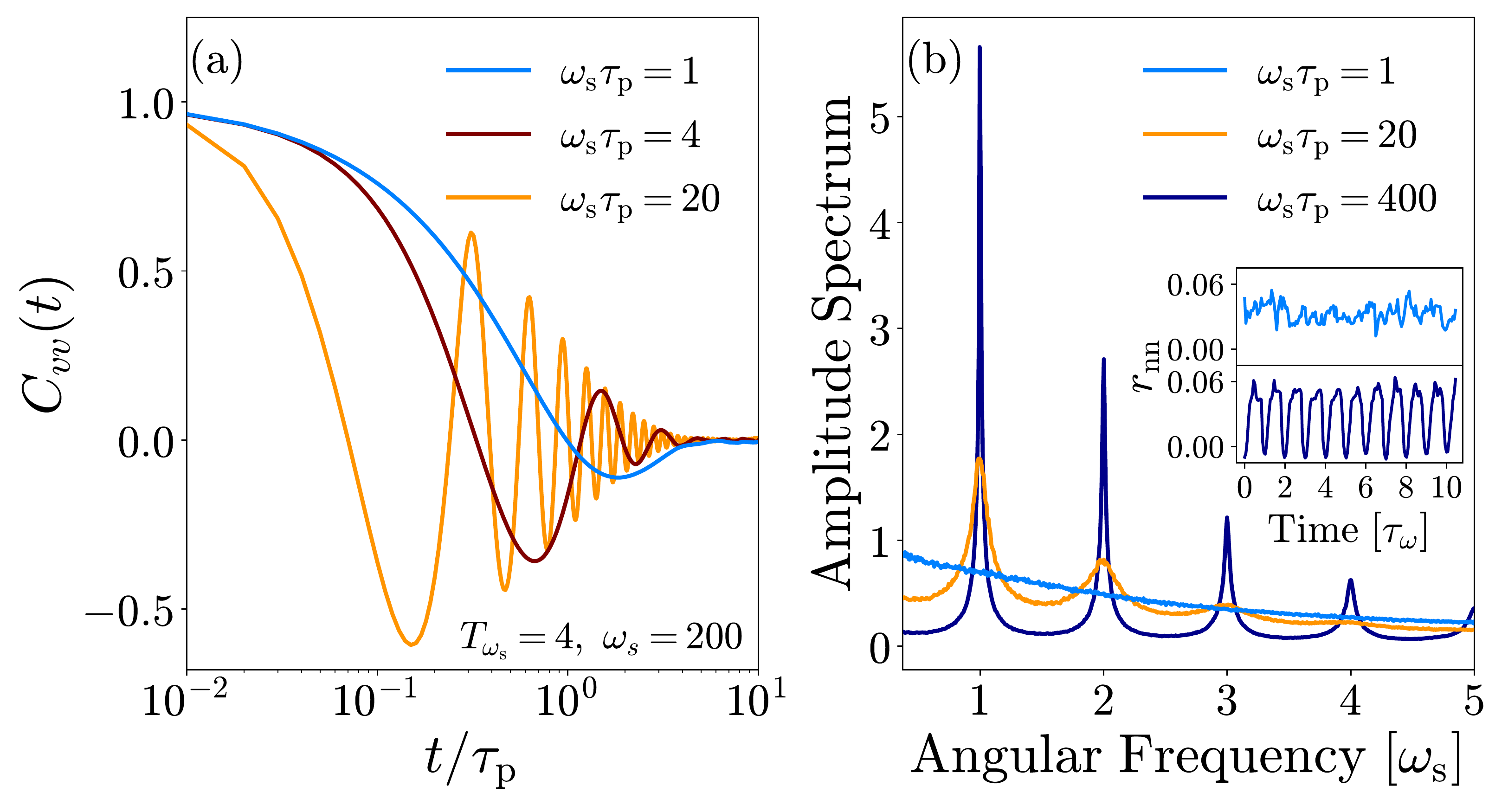} 
    \caption{(a) The velocity autocorrelation function $C_{vv}(t)$ as a function of time $t$ at $\omega_{\mathrm{s}}=200$. As $\omega_{\mathrm{s}}\tau_{\mathrm{p}}>1$ we see the emergence of oscillations which correspond to more circular particle motion inside the cage. (b) Amplitude spectrum of the nearest-neighbor distance $r_{\mathrm{nn}}(t)$ for several values of $\omega_{\mathrm{s}}\tau_{\mathrm{p}}$. Starting from  $\omega_{\mathrm{s}}\tau_{\mathrm{p}}\sim 20$, i.e., the minimum in~\cref{Fig2}, the spectrum begins to peak at $\omega_{\mathrm{s}}$ (and its multiples). This corresponds to particles periodically colliding with the same particle inside their cage. The appearance of this 'hammering' effect is especially visible in the inset where $r_{\mathrm{nn}}(t)$ for two example particles at $\omega_{\mathrm{s}}\tau_{\mathrm{p}}=1\ \mathrm{and}\ 400$ are plotted.}
    \label{Fig4}
\end{figure}

In order to verify and explain the mechanism in more detail, we will now exclusively focus on the data obtained for $\omega_{\mathrm{s}}=200$, $T_{\omega_{\mathrm{s}}}=4$ (the red squares in \cref{Fig2}), where the 'hammering' effect is strongest. We start by introducing the spinning radius $R=v_{0}/\omega_{\mathrm{s}}$ which in this case is smaller than a particle radius, i.e., $R=0.2$. If the persistence time is then also larger than the spinning period $\tau_{\omega}=2\pi/\omega_{\mathrm{s}}$ ($\tau_{\mathrm{p}}>\tau_{\omega}$ or $\omega_{\mathrm{s}}\tau_{\mathrm{p}}>2\pi$), particles should be able to undergo full circular/elliptical motion inside their cage. This can be clearly seen when inspecting multiple short-time particle trajectories (see~\cref{Fig1}b and Fig.~S7). To also quantify the periodic motion, we have extracted the normalized velocity autocorrelation function $C_{vv}(t)=\avg{\dot{\mv{r}}_{i}(0)\cdot \dot{\mv{r}}_{i}(t)}/\avg{\dot{\mv{r}}_{i}^{2}}$ for a subset of $\omega_{\mathrm{s}}\tau_{\mathrm{p}}$ values and plotted them in~\cref{Fig4}a. In accordance with the more circular trajectories, we observe the emergence of oscillations, which roughly start when $\omega_{\mathrm{s}}\tau_{\mathrm{p}}\gtrsim2\pi$ and become longer lived for increasing persistence.

Having established that particles increasingly go in circles inside their cage, we also want to show that this is indeed accompanied by systematic collisions with one (or multiple) of its surrounding particles. For this reason we have, based on short-time trajectories, extracted the nearest neighbor distance of individual particles $r_{\mathrm{nn}}(t)$ as a function of time. Two of these have been plotted as an inset in \cref{Fig4}b and confirm the notion of repeated collisions. Specifically, for $\omega_{\mathrm{s}}\tau_{\mathrm{p}}=1$, i.e., $\tau_{\mathrm{p}}<\tau_{\omega}$, the nearest neighbor distance appears to be completely random, that is, the particle undergoes random collisions with all neighboring particles. For $\omega_{\mathrm{s}}\tau_{\mathrm{p}}=400$, i.e., $\tau_{\mathrm{p}}\gg \tau_{\omega}$, the nearest neighbor distance is instead very periodic indicating that the particle collides, moves away, collides again, and so on. We have verified that these collisions are with the same neighboring particle. To check that this behavior occurs throughout our material we have also calculated the spectrum of $r_{\mathrm{nn}}(t)$ for all particles (see \cref{Fig4}b). We see that, starting from  $\omega_{\mathrm{s}}\tau_{\mathrm{p}}\sim 20$ (the minimum in~\cref{Fig2}) the spectrum begins to peak at $\omega_{\mathrm{s}}$ (and its multiples), thus further corroborating the idea of particles periodically colliding.

Overall, these results show that a 'hammering' mechanism is indeed present in our active chiral fluid and is likely to be responsible for the observed enhanced dynamics in regime II. Moreover, it also explains why the minimum of the dynamics is roughly at $\tau_{\mathrm{p}}\sim 2\tau_{\omega}$ since only from this point onward are particles able to, on average, make multiple systematic collisions with the same neighboring particle and start capitalizing on the 'hammering' effect.

%Motivate use of single particle result via the RFOT results

\textit{Conclusion.---} To conclude, our work demonstrates that chiral glassy fluids exhibit a remarkably rich dynamical phenomenology (especially when contrasted with their nonchiral counterpart~\cite{DijkstraABP2013,Voigtmann2017}) that can be characterized by a nonmonotonic (I), reentrant (II), and large persistence (III) regime. We have shown how the initial behavior (I) is fully explained by treating the surroundings of a particle as a harmonic trap and considering cage hopping of a single cABP between such traps. In the limit of extremely weak rotational diffusion (III), we have observed either speeding up or slowing down of the dynamics which is related to the (in)ability of particles to align their respective velocities and induce collective swirling-like motion. Finally, to rationalize the surprising (but for large spinning frequencies highly significant) reentrant behavior (II), we have introduced and demonstrated the existence of a new 'hammering' mechanism that is distinct for rapidly spinning particles at high densities. Overall, our results pay testimony to the fact that chirality (already in its simplest form) gives rise to a plethora of nontrivial behavior, even in the glassy limit where interactions usually dominate dynamics. 
It would be interesting to see whether these regimes and specifically the 'hammering' mechanism can also be observed and possibly exploited in an experimental setting involving for instance active granular rotors or colloids. For the latter one might also have to consider the role of translational diffusion, which could hinder collective motion or disrupt circular motion inside the cage. Alternatively, one can think of chiral active probe particles possibly utilizing the 'hammering' effect to help extract material properties or navigate through a soft dense environment such as gels~\cite{Kurzthaler2021,Velasco2020,Gomez-Solano2016,Lozano2019,Irani2022}.

%In contrast, the underlying reason for the (in some cases significant) reentrant behavior does not readily reveal itself as it is not related to any collective motion and is present in almost all considered quantities (see \cref{Fig2}

\section*{Acknowledgments}
\noindent We thank T. Voigtmann for helpful discussions. We acknowledge the Dutch Research Council (NWO) for financial support through a START-UP grant (V.E.D. and L.M.C.J.). H.L. is supported by the  Deutsche Forschungsgemeinschaft (Grant No. LO 418/22).

\appendix

\bibliographystyle{apsrev4-1}
\bibliography{all}

% Figure legends
%%Automatically it will add the figure legends  and table legends as a list by below command

\newpage

%\listoffigures

%\newpage

%\listoftables

% Figures and Tables coding should be placed where the
% first reference in the text.
% All the Figure files should be placed same working directory,
% for example (fig_1.eps and fig_1.pdf files must be present
% in the document directory)

% closing statement, nothing below matters

\end{document}